\documentclass[prd,preprintnumbers,amsmath,amssymb]{revtex4}

\input epsf
\begin{document}
\draft

\title{Canonical Ensemble Model for the Black Hole Quantum Tunneling Radiation}
\author{Jingyi Zhang\footnote{E-mail: physicz@yahoo.cn}} \affiliation{ Center for Astrophysics,
Guangzhou University, 510006, Guangzhou, China}
\date{\today}

\begin{abstract}
In this paper, a canonical ensemble model for the black hole
quantum tunneling radiation is introduced. With this model the
probability distribution function corresponding to the emission
shell is calculated. Comparing with this function, the statistical
significance of the quantum tunneling radiation spectrum of black
holes is investigated. Moreover, by calculating the entropy of the
emission shell, a discussion about the mechanism of information
flowing out from the black hole is given too.
\\\\PACS number(s): 04.70.Dy
\end{abstract}
\maketitle
\vskip2pc
\section{Introduction}
In 2000, Parikh and Wilczek presented an approach to calculate the
emission rate at which particles tunnel across the event
horizon\cite{Parikh1,Parikh2,Parikh3}. The main idea is that if
the energy conservation is taken into account, the emission
process will be a quantum tunneling, and the barrier will be
determined by the self-gravitation interaction of the emission
particle. In order to keep the spherical symmetry of the space
time during the emission process, Parikh and Wilczek treat the
tunneling particle as a spherical shell (or emission shell). In
this way a corrected spectrum, which is accurate to a first order
approximation, is given. Their result is considered to be in
agreement with an underlying unitary theory, and support the
information conservation during the emission process of the
particles. Following this method, many static or stationary
rotating black holes were studied, and similar results were
obtained
\cite{Hemming,Medved,Alves,Zhang1,Zhang2,Liu,Wu,Zhang3,Zhang4,Zhang5,Zhang6,Zhang7,Zhang8,Zhang9,Zhang10,Zhang12,Zhang13,Zhang14,Jiang,Majhi1,Majhi2,Majhi3,Kar}.
According to the Parikh-Wilczek tunneling framework, for a
Schwarzschild black hole the corrected emission spectrum is
\begin{equation}
\Gamma\sim e^{\Delta S_{BH}}=e^{-8\pi
M\omega(1-\frac{\omega}{2M})},\label{spectrum1}
\end{equation}
where $\omega$ is the energy of the emitted particle (emission
shell), $M$ is the total mass of the black hole. If $\omega$ is so
small that we can ignore the second order term, then we have
\begin{equation}
\Gamma\sim e^{\Delta S_{BH}}\thickapprox e^{-8\pi
M\omega}=e^{-\beta \omega},\label{spectrum2}
\end{equation}
here $\beta=\frac{1}{T}=-8\pi M$. On the other hand, the spectrum
given by Hawking is\cite{Hawking1,Hawking2}
\begin{equation}
N^2_{\omega}=\frac{1}{e^{\beta \omega}-1},\label{spectrum3}
\end{equation}
where $N^2_{\omega}$ denotes the intensity of the positive energy
particle flux outside and close to the event horizon, $\omega$ is
the energy level of a single emitted particle. In the condition of
classical limit, $e^{\beta \omega}\gg 1$, the Hawking radiation
spectrum will become as follows
\begin{equation}
N^2_{\omega}=e^{-\beta \omega}.\label{spectrum4}
\end{equation}
Since $N^2_{\omega}\propto\Gamma$, some people think that
Eq.(\ref{spectrum2}) and Eq.(\ref{spectrum4}) are the same, and
therefore the Parikh-Wilczek's tunneling framework and their
tunneling radiation spectrum Eq.(\ref{spectrum1}) give a
semi-classical correction to the Hawking radiation spectrum
Eq.(\ref{spectrum3}). In fact, Eq.(\ref{spectrum1}) and
Eq.(\ref{spectrum3}) are very different. In Eq.(\ref{spectrum3}),
$\omega$ is the energy level of a single emitted particle, whereas
in Eq.(\ref{spectrum1}) it denotes the energy of an emitted
spherical shell. That is, in Parikh-Wilczek's tunneling framework,
the $\omega$ is the total energy of a composite particle which
contains a number of emitted particles and constructs a spherical
shell. Thus, Eq.(\ref{spectrum1}) and Eq.(\ref{spectrum3}) reflect
different statistical significance. Moreover, most people think
that the greatest success of the Parikh-Wilczek's tunneling
framework is the consistency with the conservation of information.
However, what is the mechanism of information flowing out the
black hole? How can an emission shell carry away information? In
this paper, we first attempt to propose a canonical ensemble model
to determine the statistical significance of the tunneling
radiation spectrum. Then, with this model we calculate the entropy
of the emission shell and discuss the mechanism of information
flowing out from the hole. We use the Planck units $c=G=\hbar$
throughout the paper.
\section{canonical ensemble model corresponding to the black hole quantum tunneling radiation}
As described in section I, in Parikh-Wilczek's tunneling
framework, the tunneling particle should be a spherical shell (or
S-wave) to keep the spherical symmetry of the space time during
the emission process. It is actually equivalent to treat a
tunneling spherical shell  as a composite particle consisting of
many particles. In this framework, the tunneling and the shrinking
of the black hole take place at the same time. The tunnelling
speed of a Spherical shell is\cite{Zhang11}
\begin{equation}
\dot{r}=\frac{1}{2}(1-\frac{2(M-\omega)}{r}),\label{r}
\end{equation}
Obviously, in the vicinity of the horizon, the speed of the
tunneling shell is very slow, infinitely close to zero, and
therefore, we can think that it reaches a thermal equilibrium and
has the same temperature with the black hole. Without loss of
generality, we postulate that this composite particle consists of
identical particles, and we can treat this spherical shell as a
thermodynamical system. We imagine that the black hole acts as a
large heat source, and together with the outgoing spherical shell
constitutes an isolated system. Therefore, we can investigate the
emission shell with the canonical ensemble theory.

Suppose we have $N$ identical black hole-emission shell systems,
which constitute a mixture ensemble. let us define the statistical
operator
\begin{equation}
\hat{\rho}=\sum_i|\psi_i\big{>}P_i\big{<}\psi_i| ,\label{ss1}
\end{equation}
where $|\psi_i\big{>}$ denotes the quantum states of the emission
shell, $P_i$ is the probability of the state $|\psi_i\big{>}$.

For a canonical ensemble system, we have
\begin{equation}
P_i=\frac{\Omega_{BH}(E-E_i)}{\Omega(E)} ,\label{ss2}
\end{equation}
where $E$ and $E_i$ denote the total energy of the isolated system
and the energy of the emission shell, respectively.
$\Omega_{BH}(E-E_i)$ is the number of microscopic states of the
black hole, and $\Omega(E)$ is that of the isolated system. Since
$E_i\ll E$, in order to obtain the statistical operator
$\hat{\rho}$ and compare it with the radiation spectrum of the
Parikh-Wilczek tunnelling framework, we expand
$\ln\Omega_{BH}(E-E_i)$ into the form of Taylor series. Namely,
\begin{equation}
\ln{\Omega_{BH}(E-E_i)}=\ln{\Omega_{BH}(E)}+(\frac{\partial\ln{\Omega_{BH}}}{\partial
E_{BH}})_{E_{BH}=E}(-E_i)+\frac{1}{2}(\frac{\partial^2\ln{\Omega_{BH}}}{\partial
E^2_{BH}})_{E_{BH}=E}(-E_i)^2+\cdot\cdot\cdot.\label{ss3}
\end{equation}
Obviously,
\begin{equation}
(\frac{\partial\ln{\Omega_{BH}}}{\partial
E_{BH}})_{E_{BH}=E}=\beta=\frac{1}{K_BT},\label{ss4}
\end{equation}
\begin{equation}
(\frac{\partial^2\ln{\Omega_{BH}}}{\partial
E^2_{BH}})_{E_{BH}=E}=\frac{\partial\beta}{\partial
E_{BH}}=-\frac{K_B\beta^2}{C_{BH}},\label{ss5}
\end{equation}
here $C_{BH}$ is the heat capacity of the black hole, $T$ is the
Hawking temperature. If we calculate the Eq.(\ref{ss3}) to the
second order approximation, we obtain
\begin{equation}
P_i=\frac{\Omega_{BH}(E-E_i)}{\Omega(E)}=\frac{1}{Z}e^{-\beta
E_i-\frac{K_B}{2C_{BH}}\beta^2E^2_i},\label{ss6}
\end{equation}
where
\begin{equation}
Z=\sum_i e^{-\beta
E_i-\frac{K_B}{2C_{BH}}\beta^2E^2_i}.\label{ss62}
\end{equation}
Therefore, we have
\begin{eqnarray}
\hat{\rho}&=&\sum_i|\psi_i\big{>}\frac{1}{Z}e^{-\beta
E_i-\frac{K_B}{2C_{BH}}\beta^2E^2_i}\big{<}\psi_i|\\&=&\frac{1}{Z}e^{-\beta
\hat{H}-\frac{K_B}{2C_{BH}}\beta^2\hat{H}^2},\label{ss22}
\end{eqnarray}
where $\hat{H}$ is the Hamiltonian operator of the canonical
system. For a Schwarzschild black hole, we have
\begin{equation}
E=M,\ \ T=\frac{1}{8\pi MK_B},\ \ E_i=\omega, \ \
C_{BH}=-\frac{K_B\beta^2}{8\pi}.\label{ss7}
\end{equation}
Substituting Eq.(\ref{ss7}) into Eq.(\ref{ss6}) we get
\begin{equation}
P_i\propto e^{-8\pi M\omega(1-\frac{\omega}{2M})}=e^{\Delta
S_{BH}},\label{ss8}
\end{equation}
and
\begin{eqnarray}
\hat{\rho}=\frac{1}{Z}e^{-8\pi MK_B
\hat{H}+4\pi\hat{H}^2},\label{ss222}
\end{eqnarray}
 That is, we obtain
\begin{equation}
P_i=\Gamma\propto e^{\Delta S_{BH}}.\label{ss9}
\end{equation}
This result tells us that if we treat the black hole together with
the emission shell as an isolated system, with the canonical
ensemble, we will obtain a probability distribution function
$P_i$, which is the same as the emission rate of a spherical
shell, $\Gamma$. However, in our canonical ensemble model,
$\omega$ denotes the total energy of the system, and the $P_i$
denotes the probability of the system staying at a macrostate with
the total energy $\omega$. Of course, if we consider the emission
shell as an identical particles system, at the first order
accuracy, we can obtain the same formula as that of
Eq.(\ref{spectrum4}). Moreover, since this emission shell contains
lots of microstates, as it is emitted from the hole, it surely
carrys away a lot of information. That is, there is an out-going
information flux near the horizon during the emission process.
Now, the most interesting question is: Is the total information of
the black hole-emission shell system conservative?
\section{entropy and information of the emission shell}
In order to discuss the information puzzle during the emission
process, we first calculate the entropy of the emission shell. Let
us define a entropy operator
\begin{equation}
\hat{S}=-K_B\ln\hat{\rho}.\label{entropy1}
\end{equation}
Then, the mean value of the entropy is
\begin{equation}
S\equiv\big<\hat{S}\big>=tr(\hat\rho\hat{S})=-K_Btr(\hat\rho\ln\hat{\rho}).\label{entropy2}
\end{equation}
The exact computation about the entropy is very difficult. An
alternative approach is treating the emission shell as an
identical particles system and calculating its entropy with the
modified brick-wall method--the thin film model
\cite{Hooft,Lixiang}. That is, we treat the emission shell as a
thin film staying near the hole.

Let us take the Schwarzschild black hole as an example, and review
the thin film model of the entropy computation. The line element
of the Schwarzschild black is
\begin{equation}
ds^2=-(1-\frac{2M}{r})dt^2+(1-\frac{2M}{r})^{-1}dr^2+r^2d\Omega^2.\label{line1}
\end{equation}
The entropy of a thin film composed of identical particles near
the hole can be expressed as an integral\cite{Lixiang}
\begin{equation}
S=\frac{8\pi^3}{45\beta^3}\int_{r_H+\epsilon}^{r_H+\epsilon+\delta}\frac{r^2}{f^2}dr,\label{linn1}
\end{equation}
where $r_H=2M$ is the location of the event horizon, and
\begin{equation}
f=1-\frac{2M}{r}.\label{f1}
\end{equation}
In Eq. (\ref{linn1}), $\epsilon$ is a cutoff and $\delta$ is the
thickness of the thin film.

Now we calculate the entropy of the emission shell staying outside
the event horizon. The thickness of the emission shell
approximately equal to $r_H(M)-r_H(M-\omega)=2\omega$. Moreover,
considering the self-gravitation of emission particle,  for a
emission shell the effective space time should
be\cite{Parikh1,Parikh2,Parikh3}
\begin{equation}
ds^2=-(1-\frac{2(M-\omega)}{r})dt^2+(1-\frac{2(M-\omega)}{r})^{-1}dr^2+r^2d\Omega^2.\label{line22}
\end{equation}
That is, when we calculate the entropy of the emission shell with
the Eqs.(\ref{linn1}) and (\ref{f1}), we should replace $M$ with
$M-\omega$, and the entropy of the emission shell should be
written as
\begin{equation}
S=\frac{8\pi^3}{45\beta^3}\int_{r_H(M-\omega)+\epsilon}^{r_H(M-\omega)+\epsilon+2\omega}\frac{r^4}{(r-2M+2\omega)^2}dr,\label{S3}
\end{equation}
where $r_H(M-\omega)$ is the radial coordinate of the event
horizon, $\epsilon$ is the coordinate distance between the
emission shell and the horizon. By using the theorem of mean value
we obtain
\begin{equation}
S=\frac{8\pi^3}{45\beta^3}
\frac{r_\xi^4}{(\omega+\epsilon)^2}2\omega,\label{S4}
\end{equation}
where
\begin{equation}
r_H(M-\omega)+\epsilon<r_\xi<r_H(M-\omega)+\epsilon+2\omega.
\label{S42}
\end{equation}
In general, $\epsilon\gg\delta=2\omega$, $r_\xi\approx2M$.
Substituting $\beta=\frac{1}{T}=8\pi M$ and $r_\xi\approx2M$ into
Eq.(\ref{S4}), we have
\begin{equation}
S\approx\frac{1}{720\pi\epsilon^2}8\pi
M\omega(1-\frac{\omega}{2M}+\frac{\epsilon}{2M}).\label{S5}
\end{equation}
In the following, We discuss Eq.(\ref{S5}) for three cases:

1) If we let $\epsilon=\frac{1}{\sqrt{720\pi}}$, then we obtain
\begin{eqnarray}
S&\approx&8\pi
M\omega(1-\frac{\omega}{2M})+\sqrt{\frac{\pi}{45}}\omega\\&=&\Delta
S_{BH}+\sqrt{\frac{\pi}{45}}\omega.\label{S6}
\end{eqnarray}
The total entropy of the black hole-emission shell system is
\begin{equation}
S_{total}=S_{BH}+S=\frac{A_H(M)}{4}+\sqrt{\frac{\pi}{45}}\omega\ne
\frac{A_H(M)}{4}.\label{o}
\end{equation}
It means that the total entropy of the black hole-emission shell
system is greater than that of the black hole before emitting the
shell. In fact, according to the definition of
information\cite{Page}
\begin{equation}
I=S_{max}-S_{total},\label{I}
\end{equation}
the information $I$ decrease after emitting a shell. It means that
information lose during the emission of the black hole.

2) If $\epsilon<\frac{1}{\sqrt{720\pi}}$, then we have
\begin{eqnarray}
S>\Delta S_{BH}+\sqrt{\frac{\pi}{45}}\omega,\label{S61}
\end{eqnarray}
\begin{equation}
S_{total}=S_{BH}+S>\frac{A_H(M)}{4}+\sqrt{\frac{\pi}{45}}\omega\ne
\frac{A_H(M)}{4}.\label{1}
\end{equation}
It also means that the information $I$ decrease.

3) If $\epsilon>\frac{1}{\sqrt{720\pi}}$, the shell is far away
from the event horizon. In this case, the velocity of the shell
can not be ignored, and the thermal equilibrium with the horizon
does not exist, and therefore, the above method of calculating the
entropy will no longer adapt.
\section{conclusion}
We have introduced a canonical ensemble model corresponding to the
Parikh-Wilczek's tunnelling framework. With this model, we can
discuss not only the statistical significance of the quantum
tunneling radiation spectrum but also the mechanism of information
flowing out from the hole. We showed that the probability
distribution function for the canonical ensemble model is the same
as the tunnelling rate of the emission particles. It means that
the quantum tunnelling rate is, in fact, equal to a probability
that the black hole transits from one quantum state corresponding
to the mass $M$ to another state corresponding to the mass
$M-\omega$. We found that the emission shell contains information
and when it is emitted out from the hole, there are information
flowing out. However, according to the canonical ensemble model,
the total information is not conservative. That is, information
lose in the process of emission. It should be mentioned that our
canonical ensemble model is different from the model presented in
Ref.\cite{Zhao}. In Ref.\cite{Zhao} the authors treated the black
hole as a canonical ensemble composed of a naked black hole and
the two-dimensional thermodynamic surface (horizon of the black
hole). By using the quantum statistical method, they also derived
the energy spectrum of the black hole Hawking radiation.

 \acknowledgments This research is supported by the
National Natural Science Foundation of China (Grant Nos. 11273009,
10873003, 10573005, 10633010) and the National Basic Research
Program of China (Grant No. 2007CB815405).


\begin{references}
\bibitem{Parikh1} M. K. Parikh, F. Wilczek, \textit{Phys. Rev. Lett.},
{\bf 85}, 5042(2000) [arxiv: hep-th/9907001].
\bibitem{Parikh2} M. K. Parikh, \textit{Int. J. Mod. Phys.} D {\bf 13},2355(2004) [arXiv: hep-th/0405160].
\bibitem{Parikh3} M. K. Parikh, arXiv: hep-th/0402166,
\bibitem{Hemming} S. Hemming, E. Keski-Vakkuri, \textit{Phys. Rev.}{\bf D64},
044006(2001).
\bibitem{Medved} A. J. M. Medved,  \textit{Phys.
Rev.}{\bf D66}, 124009(2002).
\bibitem{Alves} M. A.ves, \textit{Int. J. Mod. Phys.} D {\bf10},
575(2001).
\bibitem{Zhang1} J. Zhang, Z. Zhao, \textit{Mod. Phys. Lett.} A {\bf 20}, 1673(2005).
\bibitem{Zhang2} J. Zhang, Z. Zhao, \textit{Phys. Lett.} B {\bf 618}, 14(2005).
\bibitem{Liu} W. B. Liu, \textit{Phys. Lett.} B {\bf 634},541(2006).
\bibitem{Wu} S. Q. Wu, Q. Q. Jiang, \textit{J. High Energy Phys.} (3)(2006)Art. No.079.
\bibitem{Zhang3} J. Zhang, Z. Zhao, \textit{Nucl. Phys.} B {\bf 725}, 173(2005).
\bibitem{Zhang4} J. Zhang, Z. Zhao, \textit{J. High Energy Phys.} (10)(2005)Art. No.055.
\bibitem{Zhang5} J. Zhang, Z. Zhao, \textit{Phys. Lett.} B {\bf 638}, 110(2006).
\bibitem{Zhang6} J. Zhang, Z. Zhao, \textit{Acta Phys. Sin.(in chinese)} {\bf 55}, 3796(2006).
\bibitem{Zhang7} J. Zhang, Z. Zhao, \textit{Mod. Phys. Lett.} A {\bf 21}, 1865(2006).
\bibitem{Zhang8} J. Zhang, Jun-Hui Fan, \textit{Chinese Physics}  {\bf 16}, 3879(2007).
\bibitem{Zhang9} J. Zhang, Jun-Hui Fan, \textit{Phys. Lett.} B {\bf 648}, 133(2007).
\bibitem{Zhang10} J. Zhang, \textit{Mod. Phys. Lett.} A {\bf 22}, 1821(2007).
\bibitem{Zhang12} J. Zhang, \textit{Phys. Lett.} B {\bf 668}, 353(2008).
\bibitem{Zhang13} J. Zhang, \textit{Phys. Lett.} B {\bf 675}, 14(2009).
\bibitem{Zhang14} J. Zhang, Z. Zhao, \textit{Sci. China-phys. Mech. Astron.}  {\bf 53}, 1427(2010).
\bibitem {Jiang}Q. Q. Jiang, S. Q. Wu, and X. Cai, \textit{Phys. Rev.}
\textbf{D73}, 064003(2006).
\bibitem{Majhi1} R. Banerjee, B. R. Majhi, \textit{Phys. Lett.} B {\bf 662}, 62(2008).
\bibitem{Majhi2} R. Banerjee, B. R. Majhi, S. Samanta, arXiv: 0801.3583.
\bibitem{Majhi3} R. Banerjee, B. R. Majhi, arXiv: 0805.2220.
\bibitem{Kar} S. Kar, \textit{Phys. Rev.} D {\bf 74}, 126002(2006)[arxiv: hep-th/0607029].
\bibitem{Hawking1} S. W. Hawking, \textit{Nature (London)}{\bf 248}, 30(1974).
\bibitem{Hawking2} S. W. Hawking, \textit{Commun. Math. Phys.}{\bf 43}, 199(1975).
\bibitem{Zhang11} J. Zhang, Z. Zhao, \textit{Phys. Rev.}{\bf D 83}, 064028(2011).
\bibitem{Hooft} G. t' Hooft, \textit{Nucl. Phys.} B {\bf 256}, 727(1985).
\bibitem{Lixiang} X. Li, Z. Zhao, \textit{Phys. Rev.}{\bf D 62}, 104001(2000).
\bibitem{Page} D. N. Page, hep-th/9305040v5.
\bibitem{Zhao} R. Zhao, Li-Chun Zhang, Shuang-Qi Hu, \textit{Acta Phys. Sin.(in chinese)} {\bf 55}, 3898(2006).



\end{references}
\end{document}